\pdfoutput=1

\documentclass[11pt]{article}

\usepackage[final]{acl}

\usepackage{times}
\usepackage{latexsym}
\usepackage{enumitem}
\usepackage{array}
\usepackage{booktabs}
\usepackage{float}
\usepackage[T1]{fontenc}

\usepackage[utf8]{inputenc}
\usepackage{soul}

\usepackage{microtype}

\usepackage{inconsolata}

\usepackage{graphicx}
\usepackage[table]{xcolor}
\usepackage{xcolor}
\usepackage{soul}

\title{Towards Breaking the Learning System Wall Using Multimodal Tutoring Transcriptions}

\author{
  Danielle R. Thomas \\
  Carnegie Mellon University \\
  Pittsburgh, PA, USA \\
  \texttt{drthomas@cmu.edu} \And
  Marie Cynthia Abijuru Kamikazi \\
  Carnegie Mellon University \\
  Pittsburgh, PA, USA \\
  \texttt{mabijuru@andrew.cmu.edu} \And
  Ashish Gurung \\
  Carnegie Mellon University \\
  Pittsburgh, PA, USA \\
  \texttt{agurung@andrew.cmu.edu} \AND
  Ishan Miglani \\
  Carnegie Mellon University \\
  Pittsburgh, PA, USA \\
  \texttt{imiglani@andrew.cmu.edu} \And
  Shivang Gupta \\
  Carnegie Mellon University \\
  Pittsburgh, PA, USA \\
  \texttt{shivang@cmu.edu} \And
  Zachary Levonian \\
  Renaissance Philanthropy \\
  New York, NY, USA \\
  \texttt{zach@renphil.org} \AND
  Conrad Borchers \\
  Vanderbilt University \\
  Nashville, TN, USA \\
  \texttt{c.borchers@vanderbilt.edu} \And 
  Kenneth R. Koedinger \\
  Carnegie Mellon University \\
  Pittsburgh, PA, USA \\
  \texttt{koedinger@cmu.edu}
}

\begin{document}
\maketitle
\begin{abstract}
Past research using log data has faced the ``learning system wall,'' whereby few methods exist for generalizing models of student learning across platforms. Increasingly, online learning is captured by richer forms of data, including dialog and video, with new affordances. An example of this is remote tutoring programs, where human tutors support students who use learning systems while video conferencing. Toward better platform-general modeling of learning, we introduce an AI-driven multimodal transcription system that processes screen-recording videos into unified screenplay-style transcripts containing audio dialogue and annotated learning log actions. We describe a planned method for temporally aligning AI-generated multimodal transcripts with MATHia learning logs and for identifying and classifying student learning processes to align with MATHia logs. Lastly, we highlight challenges and potential solutions in capturing learning processes in one system, offering initial steps towards generalizing log data across diverse systems.    

\end{abstract}

\section{Introduction}

Tutoring is one of the most effective educational interventions available to students \cite{nickow2020impressive}, yet relatively little is known about the specific instructional interactions that make tutoring effective \cite{chi2001learning}. To address this gap, large-scale initiatives such as the National Tutoring Observatory (NTO), SeerNet, and SafeInsights are collecting large corpora of tutoring interactions to better understand learning processes and effective instruction \cite{kizilcec2026million,manai2024connecting,zhang2026case}. 

Although math learning software can guide students through step-by-step problem solving, there is a lack of generalizability in log data between systems, referred to as the ``learning system wall''~\cite{baker2019challenges}. For example, log data from popular math learning platforms such as Carnegie Learning’s MATHia, Curriculum Associates' i-Ready, and IXL  capture and define constructs differently \cite{carnegie_learning_mathia_nd,curriculum_associates_nd,ixl_learning_nd}. For instance, what constitutes an ``attempt'' by a student or a ``hint'' generated by the system varies. A specific student may use one learning system one year and another one the next, making it difficult to monitor learning trajectories from one year to the next. This lack of generalizability poses a huge challenge for large-scale infrastructure projects aiming to develop AI models of learning, and thus improve the quality of AI-driven tutors and learning systems. 

The field of measurement and learner modeling has long grappled with challenges imposed by differences in log data format and availability across learning systems \cite{baker2021towards}. Using video data to create unified representations of problem solving offers a promising way to break this barrier. Eventually, this could allow our field to capture student actions across diverse learning platforms.

\begin{figure*}[t]
\centering
\includegraphics[width=\textwidth]{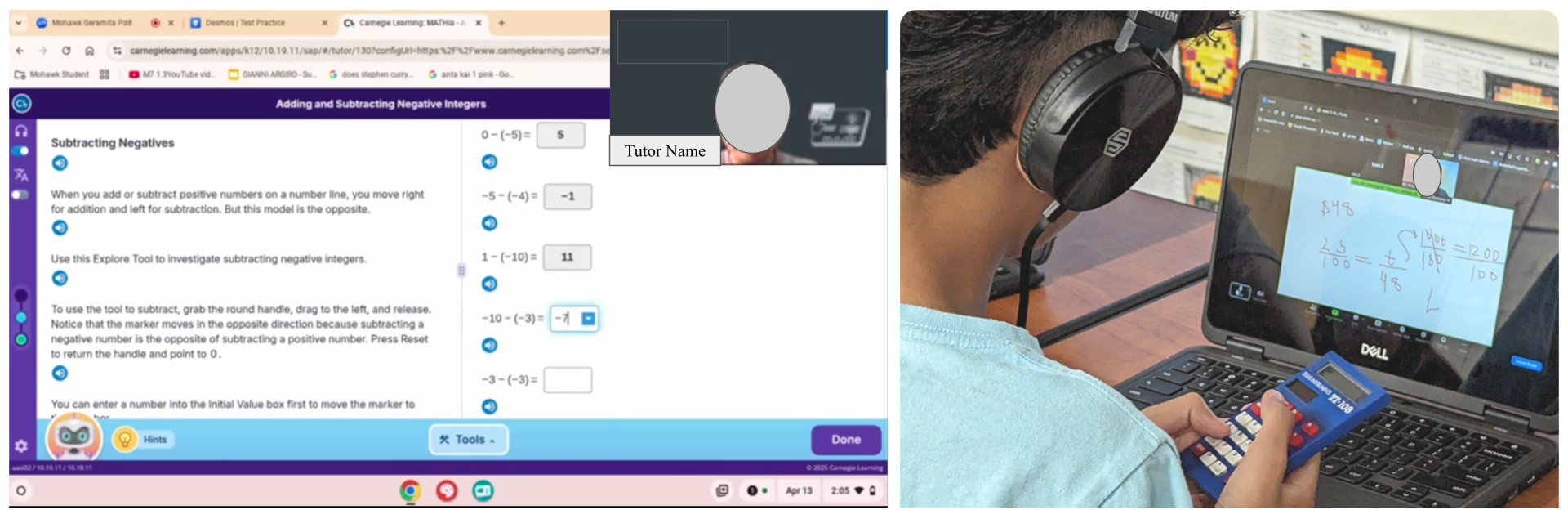}
\caption{Left: A screenshot of a student screen-share recording during remote tutoring. The student is engaging with the MATHia learning platform with the support of a remote human tutor. Right: A student is using an interactive whiteboard while receiving support from a remote human tutor.}
\label{fig:tutoring_screenshot}
\end{figure*}

Understanding learning processes is particularly important in remote human and hybrid human-AI tutoring, in which students work with a virtual human tutor while using a learning system \cite{gurung2026improving,thomas2024improving}. By combining the scalability of ITS with the personalized support of human tutors, this approach has been associated with improvements in student learning and engagement \cite{chine2022educational,gurung2025human, thomas2024improving}. Figure~\ref{fig:tutoring_screenshot} shows a screenshot of a student screen-sharing recording while engaged in remote tutoring using MATHia. 


Recent multimodal large language models (MLLMs) offer a potential method of attending to the ``learning system wall'' by generating multimodal transcripts containing log data directly from video recordings, regardless of the learning system being used \cite{yin2024survey}. Rather than relying on the alignment of separate data streams, these systems can capture both dialogue and visible on-screen actions from a single source \cite{fu2025video}. If reliable, such transcripts could provide a practical way to document student problem-solving activity across diverse learning environments, including settings where detailed log data are unavailable. 


Our contributions are two-fold: \textbf{1)} We introduce an AI-generated multimodal transcript system that captures dialogue and on-screen actions within a unified screenplay-style representation of a tutoring session; and \textbf{2)} We describe a planned method for temporally aligning multimodal transcripts with MATHia learning logs and for identifying and classifying student learning processes in multimodal transcripts to match MATHia logs. The goal of this work is to describe initial strides in a single system with the future aim of cross-system generalizability. By distilling videos into log data on a single learning platform (MATHia), we explore a method that could potentially be applied to other learning platforms, such as IXL or Curriculum Associates i-Ready. 


\section{Related Work}

\subsection{\textbf{Remote and Hybrid Human-AI Tutoring}}

Remote human tutoring involves a student working with a human tutor virtually. Hybrid human-AI tutoring combines computer-based instruction with human mentoring to address both learning and motivation \cite{thomas2024improving}. In this model, adaptive software (e.g., MATHia) guides students through practice and skill-building, while a human tutor focuses on things the software cannot, like building confidence and a sense of belonging \cite{{chine2022educational},{thomas2024improving}}. As students work on their own, tutors watch their progress in real time and step in to provide support and build rapport \cite{thomas2024improving}. This combination of human and AI support has been shown to improve student learning and engagement \cite{{borchers2026brief},{gurung2025human},{thomas2024improving}}.


\subsection{Multimodal Transcripts of Tutoring}
Historically, researchers evaluated tutoring analytics by looking primarily at isolated software logs \cite{sales2021student,he2025systematic}. ITS like MATHia can record detailed, transaction-level records of student problem-solving \cite{d2023intelligent}.  However, these data lack the real-world context of a physical classroom or video conferencing, such as student explanations, off-screen drawings, or non-verbal engagement cues \cite{borchers2026brief}. Multimodal learning analytics has emerged to integrate video, audio, and software logs into a unified representation of student learning \cite{blikstein2016multimodal}. Recently, researchers successfully aligned minute-level ITS transaction logs with Zoom audio transcripts to demonstrate how brief, real-time human tutoring visits directly counteract engagement declines \cite{borchers2026brief}. However, this multi-stream alignment process is logistically complex, frequently suffering from low match rates due to missing metadata, server clock drifts, or mismatched student Zoom aliases \cite{borchers2026brief}. This highlights a critical need for unified, automated video transcription systems that can capture both dialogue and on-screen problem solving directly from a single screen-recording source, without relying on the tedious merging of separate datastreams \cite{cukurova2020promise}.



\subsection{Challenges Facing AI-Generated Multimodal Transcription }

Using MLLMs to generate screenplay-style tutoring transcripts introduces several key challenges regarding temporal reasoning and chronological accuracy \cite{rasekh2026enhancing}. First, video-LLMs lack the sub-second temporal resolution of single-modality tools like WhisperX \cite{bain2023whisperx}, resulting in temporal drift as abstract on-screen actions are mapped alongside speech \cite{guo2025vtg}. Second, because LLMs generate text autoregressively, they struggle to serialize concurrent events from different channels (such as simultaneous talking and writing), which can missequence events and undermine micro-timing analyses \cite{schneider2024stepping}. Third, privacy regulations restrict direct human validation of raw tutoring videos that contain sensitive student data. Finally, model context window constraints require partitioning long recordings into shorter segments, which introduces boundary-related information loss and propagates alignment errors when reassembling the global timeline.

We address these challenges by: (1) partitioning long videos into overlapping segments and programmatically resolving temporal mismatches; (2) focusing strictly on on-screen visual activity; and (3) developing a method of identifying and classifying learning events that can be directly mapped to MATHia logs as a method of validating on-screen problem solving. This approach provides an objective ground truth while preserving student privacy without requiring manual video coding.

\section{\textbf{Method}}

\subsection{\textbf{Transcript Generation}}

\begin{figure}[htbp]
\includegraphics[width=0.48
\textwidth]{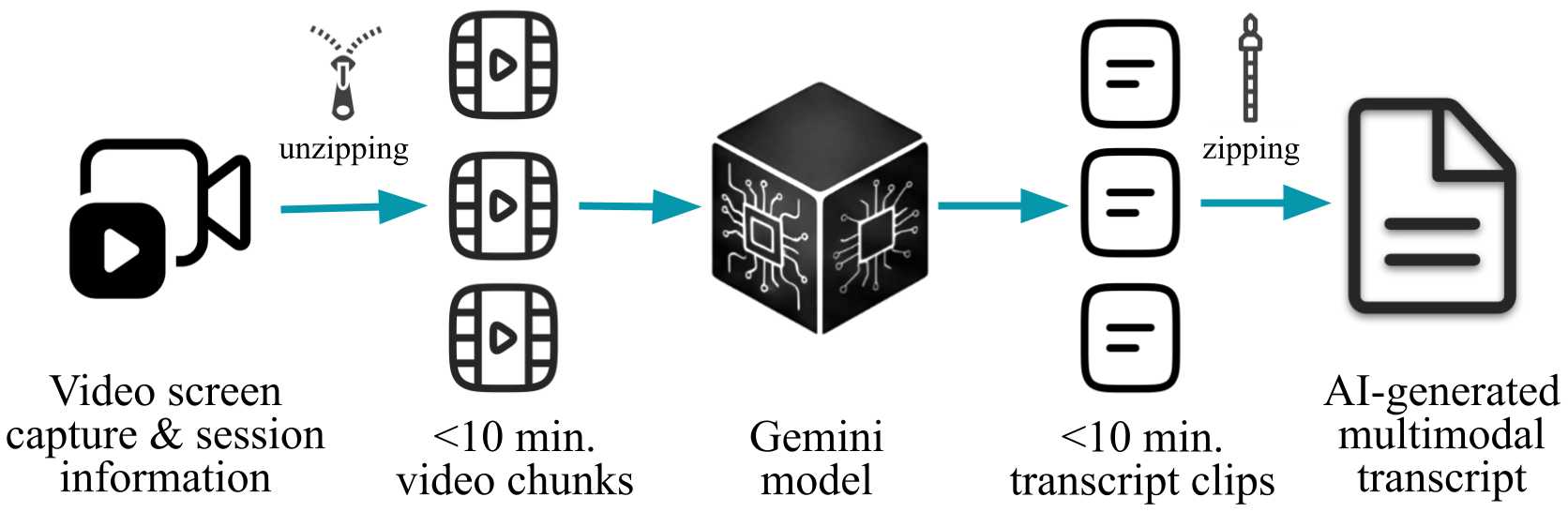}
\caption{AI-generated multimodal transcript development pipeline.}
\label{fig:pipeline}
\end{figure}

AI-generated multimodal transcription is managed by a custom automated processing architecture. Figure~\ref{fig:pipeline} illustrates the process that begins with retrieving raw Zoom videos of screen-capture recordings and their associated metadata. The system extracts rich session metadata by parsing the standardized naming convention of the source directory, which typically follows a structured string pattern: \{school\}\_\{teacher\}\_\{date\}\_\{time\}\_\{tutor\}. Through this parsing, the system extracts discrete session parameters, including the school name, classroom teacher, date, time, and assigned tutor, and cross-references them with school-specific databases. This step yields a detailed contextual summary outlining localized variables, such as the specific math software being used (e.g., MATHia) and the student hardware interface (e.g., iPads or Chromebooks). Simultaneously, because processing a long, continuous video (often lasting 30 to 60 minutes) in a single LLM request is bound by model context-window limitations, the video is partitioned using stream-copying utilities. The video is split into near-even segments (capped at a maximum of 10 minutes) with a 10-second overlapping window to avoid cutting off long events. 

Following segmentation, individual video chunks are processed through the Gemini API interface (i.e., \texttt{Gemini-2.5-pro}, \texttt{Gemini-3.5-flash}). To maximize transcription accuracy, the generated contextual summary from the metadata parser is injected directly into the model's system prompt. This metadata gives the model contextual information that helps it recognize specific tutors, classrooms, and teachers. A secondary structured prompting template enforces strict transcription constraints: verbatim dialogue must be transcribed without paraphrasing. Additionally, screen actions, including mouse cursor movements, typing, drawing, application switching, and page scrolling, must be documented in moment-by-moment detail. Periods of inactivity (such as silences while a student reads a problem) must be explicitly noted. To enforce structural consistency, the generated events are restricted to a predefined JSON output schema. This schema treats the transcript as a chronological list of discrete events with standard metadata like event offsets and durations.

Spoken dialogue is captured via \texttt{utterance events}, and on-screen activity via \texttt{mouse click} and \texttt{keyboard type events}. We aim to identify and classify on-screen problem-solving activities and map them to four MATHia-aligned event categories. Once all segment transcripts are returned, a post-processing step deduplicates redundant events in the overlapping window, and remaps the localized segment offsets back onto the global session timeline to produce a single, unified, screenplay-style multimodal transcript.

\begin{figure}[htbp]
\includegraphics[width=0.48
\textwidth]{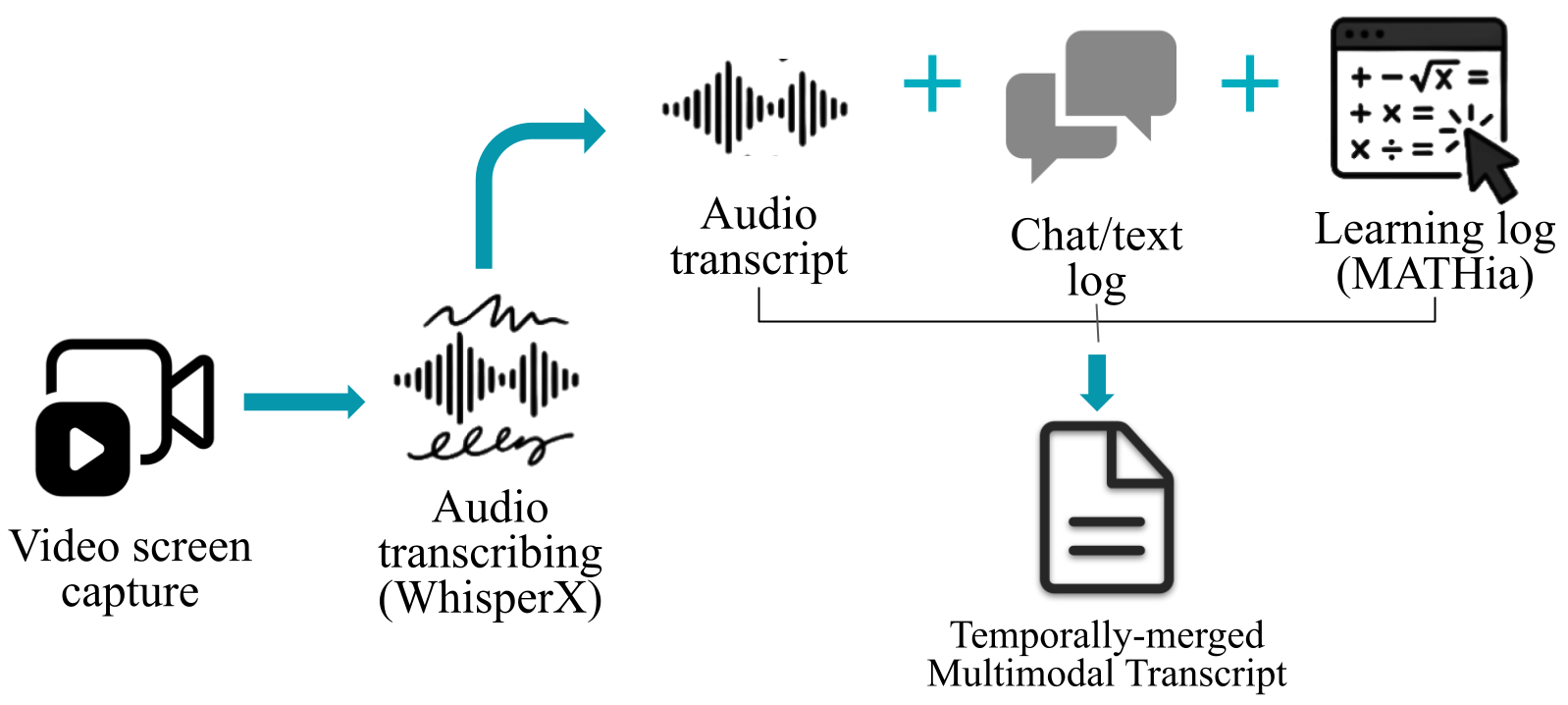}
\caption{Temporally-merged multimodal transcript (audio, chat, and MATHia logs) development pipeline.}
\label{fig:temporal_pipeline}
\end{figure}

These AI-generated multimodal transcripts can capture tutor-student interactions and student math actions that other available data streams typically do not capture, such as a student opening another tab on the screen and using supportive apps, like online graphing tools and calculators (e.g., Desmos) or a tutor and student collaboratively using an interactive whiteboard to solve a step-by-step math problem. In contrast, traditional stream-merging pipelines (Figure~\ref{fig:temporal_pipeline}) rely on aligning discrete, pre-existing datasets, such as audio transcripts, text chat history, and native ITS clickstream logs. Although this merged approach can successfully align spoken dialogue with system clicks, it remains structurally blind to any educational event that does not generate an active software log or an audible verbalization.

Traditional streaming pipelines, merging audio transcriptions, chat, and learning logs cannot capture key learning moments between a human tutor and a student, such as non-verbal learning moments or when a student works on an interactive digital whiteboard. For instance, if a student silently sketches a math model or makes an arithmetic error while drawing an equation on an open whiteboard, this action leaves no footprint in the learning log. As shown in Table~\ref{tab:transcription-example}, the traditional merged pipeline (right column) fails to capture these silent drawing events, leaving large contextual gaps in the session history. Conversely, by bypassing multi-stream alignment and directly transcribing the visual video stream (as shown in Fig. 2), our system documents these spatial-temporal actions. This direct visual transcription helps preserve actions such as a student's on-screen math errors, contextualized alongside the tutor's dialogue. 




\begin{table*}[t] 
\centering
\caption{An example of AI-generated multimodal transcription (left) and a corresponding temporally-merged multimodal transcription of audio, chat, and MATHia log data (right). AI-generated transcriptions can capture pedagogical and mathematical events not captured in available data streams, such as a student using an interactive whiteboard outside of the math learning system (its absence is shown in grey boxes.)}
\label{tab:transcription-example}
\footnotesize
\renewcommand{\arraystretch}{1.2}
\setlength{\tabcolsep}{6pt} 

\begin{tabular}{|p{1.6cm}|p{6.5cm}|p{6.5cm}|}
\hline
\textbf{Time} &
\textbf{AI-generated multimodal transcript (Fig.~\ref{fig:pipeline})} &
\textbf{Temporally-merged transcript (Fig.~\ref{fig:temporal_pipeline})} \\
\hline
13:21--13:27 & \textbf{Tutor}: I think it will be easy for me to explain. So, yeah, I think, okay. & \textbf{Tutor}: Your screen might stop sharing, because I think it would be easy for me to explain. \\
\hline
13:28--14:11 & \textbf{[view change event:} The tutor's whiteboard appears on the screen.] & \cellcolor{gray!20} \\
\hline
14:12--14:30 & \textbf{[drawing event:} tool=pen, Tutor writes the equation `74 = $x$ + $x$ - 10' on the whiteboard.] & \cellcolor{gray!20} \\
\hline
14:33--14:39 & \textbf{Tutor}: Now we have a simple equation and we just need to find the value of $x$. & \textbf{Tutor:} Now we have a simple equation and we just need to find the value of $x$. \\
\hline
14:41--14:48 & \textbf{Tutor}: Um, I think even you can write. So if you want to write on the whiteboard itself, you can feel free to go ahead and write as well. & \textbf{Tutor}: I think even you can write. So if you want to write on the whiteboard itself, you can be free to go ahead and write as well. \\
\hline
15:44--15:47 & \textbf{[drawing event:} tool=pen, The student starts to write `84'.] & \cellcolor{gray!20} \\ 
\hline
15:48--15:51 & \textbf{Tutor:} 84 which is equal to $2x$ & \textbf{Tutor:} Which is equal to till $x$ \\
\hline
15:53--15:54 & \textbf{Tutor:} So what do you think is the value of $x$? & \textbf{Tutor:}  So what do you think is the value of $x$?\\
\hline
15:54--15:58 & \textbf{[drawing event:} tool=pen, The student writes `42'.] & \cellcolor{gray!20} \\
\hline
15:59--16:04 & \textbf{Tutor}: Yeah, that's great. You solved it. & \textbf{Tutor:} Yeah, that's great. You solved it. \\
\hline
16:09--16:12 & \textbf{[keyboard type event:} text `42'; Student types `42' into the input box for the value of $x$.] & \textbf{Attempt:} {value=42, escape-in-messages=false}\\
\hline
\end{tabular}
\end{table*}

\subsection{Data collection}
To ensure stringent data privacy, all transcripts and corresponding system logs are fully de-identified and anonymized prior to analysis. The anonymization process includes the removal of personally identifiable information (PII) present in the transcripts of the students' shared screens. MATHia data is de-identified and anonymized. To prevent data leakage of sensitive video recordings, the transcript generation pipeline utilizes locally hosted MLLMs, ensuring that student data remains entirely within a secure local environment and is never used for external model training. All data collection and processing procedures are conducted in strict compliance with approved IRB protocols.

\subsection{Planned procedure for temporal alignment }
A critical challenge in multi-stream integration is temporal misalignment caused by variable startup times (e.g., tutors and students entering Zoom rooms and break-out rooms at different times), network latency (lag in learning log data recording, system logging) and clock drift (LLMs' accuracy in recording time in transcriptions, temporal shifts when merging chunks of recordings) \cite{borchers2026brief}. Here, we describe a planned procedure for temporally aligning the AI-generated multimodal transcripts and the corresponding native MATHia learning logs. 

First, we identify the first physical action by the student such as a \textit{keystroke\_event} or \textit{mouse\_click\_event} after the student has logged into MATHia while working with the human tutor and video conferencing. The first relative video timestamp is mapped to the absolute server timestamp of the first logged `Attempt' in the MATHia clickstream log. By calculating the difference between these two points, the system establishes a baseline temporal offset to project the rest of the relative video timeline onto an absolute scale.

However, if the student already logged into MATHia before the human tutor entered the breakout room for which the tutor is providing a recording, we will have to rely on detecting the first visually observable physical event (such as an on-screen mouse click, cursor movement, or keyboard entry) within the active video frame. Here the multimodal transcript may state, \textit{``View changes to a different student's ([student name]) MATHia workspace''}, \textit{``Video cuts to a black screen, then a new screen with a split view of Zoom and MATHia''}, and \textit{``Student's screen share starts, showing the MATHia interface. The problem is titled Modeling Integer Rates of Change.''} Then we map its relative timestamp directly to the timestamp of the first recorded student action in the backend MATHia log to compute the necessary temporal offset.

One concern is clock drift, which can accumulate over the course of a tutoring session (arising from hardware system clock variances, network delays, or minor inaccuracies in the MLLM's temporal extraction). Relying solely on a single initial anchor may introduce misalignment over time. To mitigate this issue, we will regularly perform manual re-calibrations by checking intermediate visual milestones in the transcriptions against distinct, later-occurring backend MATHia event logs, allowing us to assess active clock drift and realign the temporal offsets as necessary throughout the session.

\subsection{Planned procedure for identifying and classifying learning processes}
We describe a planned method for identifying and classifying student learning processes aligning with MATHia logs to include within the multimodal transcript. 
First, we will create a development set consisting of approximately 50 transcripts, where we manually align multimodal transcripts and corresponding MATHia logs. We then will use algorithmic methods to determine which words and phrases in the multimodal transcription are the most predictive of the four learning event types in MATHia. We will then apply an input/output classification script (similar to the one in Table~\ref{tab:event-schema-mapping}) to classify the descriptive text of the AI-generated transcriptions into four math log events: \textit{attempt}, \textit{hint\_request}, \textit{hint\_level\_change}, and \textit{done}. For instance, for \textit{``Student types `36' in the `Enter the value of x text box',}'' the phrase \textit{`student types'} implies an `Attempt.' Similarly, for \textit{``The hint content updates to show `Hint 3 of 3'},'' the phrase \textit{`Hint 3 of 3'} implies the student requested a third hint indicative of a `hint\_level\_change.'

To validate the reliability and robustness of this classification process, we will evaluate the classification script on a separate holdout testing set with manually-annotated tutoring transcripts that were not used during the heuristic development phase. We will calculate standard classification metrics -- precision, recall, and F1-score -- for each of the four event types, alongside Cohen's Kappa ($\kappa$) to measure the level of agreement between the automated classifier and the manual ground-truth alignments. 

Finally, we will validate the generalizability of the classification rules by comparing the performance metrics achieved on the testing set against those obtained on the development set. A minimal degradation in F1-score and Cohen's Kappa when transitioning from the development set to the unseen testing set will demonstrate that our keyword-matching heuristics and predictive phrase associations generalize effectively to new tutoring sessions without overfitting to the development data.

\begin{table*}[t] 
\centering
\caption{Input/output classification used to map AI-generated transcriptions and MATHia ground truth to event type.}
\label{tab:event-schema-mapping}
\footnotesize
\renewcommand{\arraystretch}{1.2}
\setlength{\tabcolsep}{6pt} 

\begin{tabular}{|p{2.2cm}|p{6.5cm}|p{5.8cm}|}
\hline
\textbf{Event Type} &
\textbf{Input/AI-Generated Multimodal Transcription} &
\textbf{Output/Ground Truth (MATHia)} \\
\hline
\textit{attempt} &
Visual detection of keyboard input, mouse drags, or system feedback while using MATHia containing ``type'', ``types'', ``plots'', ``drags'', ``selects'', etc. (e.g., \textit{``Student types 30 into the input box''}; \textit{``text `13' into the reflection line value box''}). \textbf{Correct}: ``green checkmark appears'', ``page advances'', ``next screen loads''; \textbf{Incorrect}: ``Try again modal appears'', ``Error message appears'', ``Orange (or red) input highlights'', ``try again'', ``incorrect'', etc. &

Action = ``Attempt'' contains input text data (e.g., \{value = x\}) and correctness data (e.g., OK or ERROR). \\
\hline
\textit{hint\_request} &
Visual detection of initial hint interaction while using MATHia containing keywords such as ``hint'' or ``hints'' (e.g., \textit{``Student clicks Hints button''}; \textit{``Hint pop-up appears, showing Hint 1 of 3''}; \textit {``The next button hint appears''}). Distinguished from hint\_level\_change by being the first event in the hint sequence. &
Action = ``Hint Request'', just-in-time ``JIT'', or outcome = ``INITIAL\_HINT''. \\
\hline
\textit{hint\_level\_change} &
Visual detection of advancing hint states (e.g., \textit{``Student clicks Next button in hint pop-up''}; \textit{``Hint pop-up changes to show Hint 2 of 3''}). &
Action = ``Hint Level Change'' or outcome = ``HINT\_LEVEL\_CHANGE''. \\
\hline
\textit{done} &
Visual detection of problem-level completion indicated by ``done'' or ``I'm done''  (e.g., \textit{``Student clicks the \texttt{I'm Done} button''}). &
Action = ``Attempt'' with input = ``Done'' or step\_name = ``Done''. \\
\hline
\end{tabular}
\end{table*}

\section{Challenges}
There are many challenges related to creating AI-generated multimodal transcripts containing native problem solving logs native to student math software such as MATHia. First, temporal matching between the video recordings and MATHia learning logs is difficult because students could enter a Zoom breakout room at any point within a video recording. In addition, some students with MATHia log data may not have AI-generated transcript data. In reality, this may be due to incomplete video recordings being uploaded to the system. Second, while the multimodal transcription system is designed to capture both visual actions and spoken dialogue, our current plan focuses exclusively on validating on-screen, math-logged events; we have not yet determined the accuracy of the system's integrated audio transcriptions. Third, we anticipate our method may log other learning process events that will not be present in a student's corresponding MATHia log, such as online calculator use (e.g., Desmos) or interactive whiteboard activity for assistance with problem solving. This present plan does not have any method of discriminating between MATHia-aligned events versus other learning processes. Fourth, addressing cumulative clock drift is challenging. When aligning different data streams (like AI-generated transcriptions and learning logs), the timelines often naturally drift apart over time as AI-generated transcriptions are not keeping time in a standardized manner like learning logs do. To keep these files in sync over long sessions, researchers have developed algorithmic methods to constantly calculate and correct this timing drift \cite{yiugitler2020overview}.  While we can mitigate drift by manually re-calibrating and realigning the relative timeline against backend MATHia logs, manual assessment and periodic adjustment limit the system's current capacity for fully automated, hands-off synchronization at scale.



\section{Discussion}
\textbf{We demonstrate a planned procedure for distilling videos into learning log data within a single case study.} Although our long-term goal is to bridge the ``learning system wall'' \cite{baker2019challenges,baker2021towards} across diverse platforms like IXL and i-Ready, we present this work as an initial case study focused on a single platform. Using native MATHia logs as ground truth, this study establishes a foundational baseline for distilling screen recordings into structured log data. This focused case study represents a key initial step toward validating cross-system generalizability, which will require further empirical testing on other platforms in future work. 

\textbf{Our AI-generated transcription system can capture non-verbal behaviors and ecological context not included in typical data streams.} This video-based approach captures critical student behaviors lost in audio-only logs (see Table~\ref{tab:transcription-example}). Visually, our system documents essential ecological context, such as a student being distracted or leaving the computer—behaviors that audio misses and native logs only label as ``inactivity'', or in the case of MATHia ``idleness'' \cite{cukurova2020promise}. 

In addition, in noisy classrooms, students often rely on silent typing to interact with remote human tutors. Relying on traditional log data obscures these interactions. For example, in one of our analyzed sessions, the native MATHia log simply recorded an erroneous attempt. However, our AI-generated transcript revealed the student was actually typing, \textit{``I do, but I don't really know how to use it''}  into the Zoom chat—capturing a cry for help that the system log entirely missed. 

Our screen-shared video transcription method can also capture tutors' and students' problem-solving process outside of MATHia, such as writing directly on the screen (see Table~\ref{tab:transcription-example}) or using an interactive whiteboard (see Figure~\ref{fig:tutoring_screenshot}, right), a student switching apps, or accessing other platforms for learning. For instance, our system explicitly captured a student switching to an external tool (\textit{``Student deletes `-2' from the expression in Desmos'' followed by ``Student clicks the division button on the Desmos calculator'' (target=division button)}). This provided crucial context for how the student calculated the complex fraction subsequently logged by MATHia; the native log merely registered the final answer, completely blind to the external tool usage. Furthermore, the transcriptions reveal major instructional shifts, such as when a tutor stops the student's screen share to start sharing a whiteboard. In this scenario, MATHia merely logged a successful attempt. Relying solely on the MATHia log would completely erase the tutor’s whiteboard scaffolding from the learning record.




Ultimately, using AI-generated multimodal transcriptions to provide a ``screenplay'' of student learning shows promise in creating generalizable systems. There are two advantages: 1) This method captures the specific math and problem-solving steps students are working on, alongside their ecological context, even in the absence of native learning log data. 2) This method eliminates the need to tediously merge disparate data streams (e.g., audio transcripts, chat logs, and clickstreams) by generating a unified chronological record of the tutoring session directly from the video.

\section{Conclusion}
In this work in progress, we present an AI-driven multimodal transcription system to address the ``learning system wall'' by converting tutoring screen recordings into screenplay-style transcripts. Rather than merging complex software logs, this approach captures critical student behaviors, such as whiteboard drawings, chat messages, and the use of external tools. To ensure alignment accuracy, we plan to synchronize these transcripts with MATHia logs, introducing a periodic manual re-calibration process to correct for cumulative clock drift. We then describe classifying transcript actions into standardized learning events. Although challenges in automation and platform-specific formatting remain, this methodology establishes an initial procedure for transcribing screen recordings into structured interaction data from other educational systems.

\section*{Acknowledgments}
This work was made possible with the support of the Learning Engineering Virtual Institute. We also thank system developers AJ Strauman-Scott and Ruth Schäfer for their significant contributions. The opinions, findings, and conclusions expressed in this material are those of the authors.

\bibliography{custom}

\end{document}